\def\unlock{\catcode`@=11} 
\def\lock{\catcode`@=12} 
\unlock
%
%
%
%
%

\font\fourteenrm=cmr10 scaled\magstep2
\font\twelverm=cmr10 scaled\magstep1
\font\ninerm=cmr9          \font\sixrm=cmr6

\font\fourteenbf=cmbx10 scaled\magstep2
\font\twelvebf=cmbx10 scaled\magstep1

\font\seventeeni=cmmi10 scaled\magstep3     \skewchar\seventeeni='177
\font\fourteeni=cmmi10 scaled\magstep2      \skewchar\fourteeni='177
\font\twelvei=cmmi10 scaled\magstep1        \skewchar\twelvei='177
\font\ninei=cmmi9                           \skewchar\ninei='177
\font\sixi=cmmi6                            \skewchar\sixi='177
\font\seventeensy=cmsy10 scaled\magstep3    \skewchar\seventeensy='60
\font\fourteensy=cmsy10 scaled\magstep2     \skewchar\fourteensy='60
\font\twelvesy=cmsy10 scaled\magstep1       \skewchar\twelvesy='60
\font\ninesy=cmsy9                          \skewchar\ninesy='60
\font\sixsy=cmsy6                           \skewchar\sixsy='60

\font\fourteenex=cmex10 scaled\magstep2
\font\twelveex=cmex10 scaled\magstep1

\font\fourteensl=cmsl10 scaled\magstep2
\font\twelvesl=cmsl10 scaled\magstep1

\font\fourteenit=cmti10 scaled\magstep2
\font\twelveit=cmti10 scaled\magstep1
\font\twelvett=cmtt10 scaled\magstep1
\font\twelvecp=cmcsc10 scaled\magstep1
\font\tencp=cmcsc10
\newfam\cpfam
%
%
\newcount\f@ntkey            \f@ntkey=0
\def\samef@nt{\relax \ifcase\f@ntkey \rm \or\oldstyle \or\or
         \or\it \or\sl \or\bf \or\tt \or\caps \fi }
\def\fourteenpoint{\relax
    \textfont0=\fourteenrm          \scriptfont0=\tenrm
    \scriptscriptfont0=\sevenrm
     \def\rm{\fam0 \fourteenrm \f@ntkey=0 }\relax
    \textfont1=\fourteeni           \scriptfont1=\teni
    \scriptscriptfont1=\seveni
     \def\oldstyle{\fam1 \fourteeni\f@ntkey=1 }\relax
    \textfont2=\fourteensy          \scriptfont2=\tensy
    \scriptscriptfont2=\sevensy
    \textfont3=\fourteenex     \scriptfont3=\fourteenex
    \scriptscriptfont3=\fourteenex
    \def\it{\fam\itfam \fourteenit\f@ntkey=4 }\textfont\itfam=\fourteenit
    \def\sl{\fam\slfam \fourteensl\f@ntkey=5 }\textfont\slfam=\fourteensl
    \scriptfont\slfam=\tensl
    \def\bf{\fam\bffam \fourteenbf\f@ntkey=6 }\textfont\bffam=\fourteenbf
    \scriptfont\bffam=\tenbf     \scriptscriptfont\bffam=\sevenbf
    \def\tt{\fam\ttfam \twelvett \f@ntkey=7 }\textfont\ttfam=\twelvett
    \h@big=11.9\p@{} \h@Big=16.1\p@{} \h@bigg=20.3\p@{} \h@Bigg=24.5\p@{}
    \def\caps{\fam\cpfam \twelvecp \f@ntkey=8 }\textfont\cpfam=\twelvecp
    \setbox\strutbox=\hbox{\vrule height 12pt depth 5pt width\z@}
    \samef@nt}
\def\twelvepoint{\relax
    \textfont0=\twelverm          \scriptfont0=\ninerm
    \scriptscriptfont0=\sixrm
     \def\rm{\fam0 \twelverm \f@ntkey=0 }\relax
    \textfont1=\twelvei           \scriptfont1=\ninei
    \scriptscriptfont1=\sixi
     \def\oldstyle{\fam1 \twelvei\f@ntkey=1 }\relax
    \textfont2=\twelvesy          \scriptfont2=\ninesy
    \scriptscriptfont2=\sixsy
    \textfont3=\twelveex          \scriptfont3=\twelveex
    \scriptscriptfont3=\twelveex
    \def\it{\fam\itfam \twelveit \f@ntkey=4 }\textfont\itfam=\twelveit
    \def\sl{\fam\slfam \twelvesl \f@ntkey=5 }\textfont\slfam=\twelvesl
    \scriptfont\slfam=\ninerm
    \def\bf{\fam\bffam \twelvebf \f@ntkey=6 }\textfont\bffam=\twelvebf
    \scriptfont\bffam=\ninerm     \scriptscriptfont\bffam=\sixrm
    \def\tt{\fam\ttfam \twelvett \f@ntkey=7 }\textfont\ttfam=\twelvett
    \h@big=10.2\p@{}
    \h@Big=13.8\p@{}
    \h@bigg=17.4\p@{}
    \h@Bigg=21.0\p@{}
    \def\caps{\fam\cpfam \twelvecp \f@ntkey=8 }\textfont\cpfam=\twelvecp
    \setbox\strutbox=\hbox{\vrule height 10pt depth 4pt width\z@}
    \samef@nt}
\def\tenpoint{\relax
    \textfont0=\tenrm          \scriptfont0=\sevenrm
    \scriptscriptfont0=\fiverm
    \def\rm{\fam0 \tenrm \f@ntkey=0 }\relax
    \textfont1=\teni           \scriptfont1=\seveni
    \scriptscriptfont1=\fivei
    \def\oldstyle{\fam1 \teni \f@ntkey=1 }\relax
    \textfont2=\tensy          \scriptfont2=\sevensy
    \scriptscriptfont2=\fivesy
    \textfont3=\tenex          \scriptfont3=\tenex
    \scriptscriptfont3=\tenex
    \def\it{\fam\itfam \tenit \f@ntkey=4 }\textfont\itfam=\tenit
    \def\sl{\fam\slfam \tensl \f@ntkey=5 }\textfont\slfam=\tensl
    \def\bf{\fam\bffam \tenbf \f@ntkey=6 }\textfont\bffam=\tenbf
    \scriptfont\bffam=\sevenbf     \scriptscriptfont\bffam=\fivebf
    \def\tt{\fam\ttfam \tentt \f@ntkey=7 }\textfont\ttfam=\tentt
    \def\caps{\fam\cpfam \tencp \f@ntkey=8 }\textfont\cpfam=\tencp
    \setbox\strutbox=\hbox{\vrule height 8.5pt depth 3.5pt width\z@}
    \samef@nt}
%
%
%
%
\newdimen\h@big  \h@big=8.5\p@
\newdimen\h@Big  \h@Big=11.5\p@
\newdimen\h@bigg  \h@bigg=14.5\p@
\newdimen\h@Bigg  \h@Bigg=17.5\p@
\def\big#1{{\hbox{$\left#1\vbox to\h@big{}\right.\n@space$}}}
\def\Big#1{{\hbox{$\left#1\vbox to\h@Big{}\right.\n@space$}}}
\def\bigg#1{{\hbox{$\left#1\vbox to\h@bigg{}\right.\n@space$}}}
\def\Bigg#1{{\hbox{$\left#1\vbox to\h@Bigg{}\right.\n@space$}}}
%
%
%
\normalbaselineskip = 20pt plus 0.2pt minus 0.1pt
\normallineskip = 1.5pt plus 0.1pt minus 0.1pt
\normallineskiplimit = 1.5pt
\newskip\normaldisplayskip
\normaldisplayskip = 18pt plus 4pt minus 8pt
\newskip\normaldispshortskip
\normaldispshortskip = 5pt plus 4pt
\newskip\normalparskip
\normalparskip = 6pt plus 2pt minus 1pt
\newskip\skipregister
\skipregister = 5pt plus 2pt minus 1.5pt
\newif\ifsingl@    \newif\ifdoubl@
\newif\iftwelv@    \twelv@true
\def\singlespace{\singl@true\doubl@false\spaces@t}
\def\doublespace{\singl@false\doubl@true\spaces@t}
\def\normalspace{\singl@false\doubl@false\spaces@t}
\def\Tenpoint{\tenpoint\twelv@false\spaces@t}
\def\Twelvepoint{\twelvepoint\twelv@true\spaces@t}
\def\spaces@t{\relax%
 \iftwelv@ \ifsingl@\subspaces@t3:4;\else\subspaces@t1:1;\fi%
 \else \ifsingl@\subspaces@t3:5;\else\subspaces@t4:5;\fi \fi%
 \ifdoubl@ \multiply\baselineskip by 5%
 \divide\baselineskip by 4 \fi \unskip}
\def\subspaces@t#1:#2;{%
      \baselineskip = \normalbaselineskip%
      \multiply\baselineskip by #1 \divide\baselineskip by #2%
      \lineskip = \normallineskip%
      \multiply\lineskip by #1 \divide\lineskip by #2%
      \lineskiplimit = \normallineskiplimit%
      \multiply\lineskiplimit by #1 \divide\lineskiplimit by #2%
      \parskip = \normalparskip%
      \multiply\parskip by #1 \divide\parskip by #2%
      \abovedisplayskip = \normaldisplayskip%
      \multiply\abovedisplayskip by #1 \divide\abovedisplayskip by #2%
      \belowdisplayskip = \abovedisplayskip%
      \abovedisplayshortskip = \normaldispshortskip%
      \multiply\abovedisplayshortskip by #1%
        \divide\abovedisplayshortskip by #2%
      \belowdisplayshortskip = \abovedisplayshortskip%
      \advance\belowdisplayshortskip by \belowdisplayskip%
      \divide\belowdisplayshortskip by 2%
      \smallskipamount = \skipregister%
      \multiply\smallskipamount by #1 \divide\smallskipamount by #2%
      \medskipamount = \smallskipamount \multiply\medskipamount by 2%
      \bigskipamount = \smallskipamount \multiply\bigskipamount by 4 }
\def\normalbaselines{ \baselineskip=\normalbaselineskip%
   \lineskip=\normallineskip \lineskiplimit=\normallineskip%
   \iftwelv@\else \multiply\baselineskip by 4 \divide\baselineskip by 5%
     \multiply\lineskiplimit by 4 \divide\lineskiplimit by 5%
     \multiply\lineskip by 4 \divide\lineskip by 5 \fi }
\Twelvepoint  
\interlinepenalty=50
\interfootnotelinepenalty=5000
\predisplaypenalty=9000
\postdisplaypenalty=500
\hfuzz=1pt
\vfuzz=0.2pt
%
%
%
\def\pagecontents{%
   \ifvoid\topins\else\unvbox\topins\vskip\skip\topins\fi
   \dimen@ = \dp255 \unvbox255
   \ifvoid\footins\else\vskip\skip\footins\footrule\unvbox\footins\fi
   \ifr@ggedbottom \kern-\dimen@ \vfil \fi }
\def\makeheadline{\vbox to 0pt{ \skip@=\topskip
      \advance\skip@ by -12pt \advance\skip@ by -2\normalbaselineskip
      \vskip\skip@ \line{\vbox to 12pt{}\the\headline} \vss
      }\nointerlineskip}
\def\makefootline{\baselineskip = 1.5\normalbaselineskip
                 \line{\the\footline}}
\newif\iffrontpage
\newif\ifletterstyle
\newif\ifp@genum
\def\nopagenumbers{\p@genumfalse}
\def\pagenumbers{\p@genumtrue}
\pagenumbers
\newtoks\paperheadline
\newtoks\letterheadline
\newtoks\letterfrontheadline
\newtoks\lettermainheadline
\newtoks\paperfootline
\newtoks\letterfootline
\newtoks\date
\footline={\ifletterstyle\the\letterfootline\else\the\paperfootline\fi}
\paperfootline={\hss\iffrontpage\else\ifp@genum\tenrm\folio\hss\fi\fi}
\letterfootline={\hfil}
\headline={\ifletterstyle\the\letterheadline\else\the\paperheadline\fi}
\paperheadline={\hfil}
\letterheadline{\iffrontpage\the\letterfrontheadline
     \else\the\lettermainheadline\fi}
\lettermainheadline={\rm\ifp@genum page \ \folio\fi\hfil\the\date}
\def\monthname{\relax\ifcase\month 0/\or January\or February\or
   March\or April\or May\or June\or July\or August\or September\or
   October\or November\or December\else\number\month/\fi}
\date={\monthname\ \number\day, \number\year}
\countdef\pagenumber=1  \pagenumber=1
\def\advancepageno{\global\advance\pageno by 1
   \ifnum\pagenumber<0 \global\advance\pagenumber by -1
    \else\global\advance\pagenumber by 1 \fi \global\frontpagefalse }
\def\folio{\ifnum\pagenumber<0 \romannumeral-\pagenumber
           \else \number\pagenumber \fi }
\def\footrule{\dimen@=\prevdepth\nointerlineskip
   \vbox to 0pt{\vskip -0.25\baselineskip \hrule width 0.35\hsize \vss}
   \prevdepth=\dimen@ }
\newtoks\foottokens
\foottokens={\Tenpoint\singlespace}
\newdimen\footindent
\footindent=24pt
\def\vfootnote#1{\insert\footins\bgroup  \the\foottokens
   \interlinepenalty=\interfootnotelinepenalty \floatingpenalty=20000
   \splittopskip=\ht\strutbox \boxmaxdepth=\dp\strutbox
   \leftskip=\footindent \rightskip=\z@skip
   \parindent=0.5\footindent \parfillskip=0pt plus 1fil
   \spaceskip=\z@skip \xspaceskip=\z@skip
   \Textindent{$ #1 $}\footstrut\futurelet\next\fo@t}
\def\Textindent#1{\noindent\llap{#1\enspace}\ignorespaces}
\def\footnote#1{\attach{#1}\vfootnote{#1}}

\let\footsymbol=\star
\newcount\lastf@@t           \lastf@@t=-1
\newcount\footsymbolcount    \footsymbolcount=0
\newif\ifPhysRev
\def\footsymbolgen{\relax \ifPhysRev \iffrontpage \NPsymbolgen\else
      \PRsymbolgen\fi \else \NPsymbolgen\fi
   \global\lastf@@t=\pageno \footsymbol }
\def\NPsymbolgen{\ifnum\footsymbolcount<0 \global\footsymbolcount=0\fi
   {\iffrontpage \else \advance\lastf@@t by 1 \fi
    \ifnum\lastf@@t<\pageno \global\footsymbolcount=0
     \else \global\advance\footsymbolcount by 1 \fi }
   \ifcase\footsymbolcount \fd@f\star\or \fd@f\dagger\or \fd@f\ddagger\or
    \fd@f\ast\or \fd@f\natural\or \fd@f\diamond\or \fd@f\bullet\or
    \fd@f\nabla\else \fd@f\dagger\global\footsymbolcount=0 \fi }
\def\fd@f#1{\xdef\footsymbol{#1}}
\def\PRsymbolgen{\ifnum\footsymbolcount>0 \global\footsymbolcount=0\fi
      \global\advance\footsymbolcount by -1
      \xdef\footsymbol{\sharp\number-\footsymbolcount} }
\def\space@ver#1{\let\@sf=\empty \ifmmode #1\else \ifhmode
   \edef\@sf{\spacefactor=\the\spacefactor}\unskip${}#1$\relax\fi\fi}
\def\attach#1{\space@ver{\strut^{\mkern 2mu #1} }\@sf\ }
%
%
%
\newcount\chapternumber      \chapternumber=0
\newcount\sectionnumber      \sectionnumber=0
\newcount\equanumber         \equanumber=0
\let\chapterlabel=0
\newtoks\chapterstyle        \chapterstyle={\Number}
\newskip\chapterskip         \chapterskip=\bigskipamount
\newskip\sectionskip         \sectionskip=\medskipamount
\newskip\headskip            \headskip=8pt plus 3pt minus 3pt
\newdimen\chapterminspace    \chapterminspace=15pc
\newdimen\sectionminspace    \sectionminspace=10pc
\newdimen\referenceminspace  \referenceminspace=25pc
\def\chapterreset{\global\advance\chapternumber by 1
   \ifnum\equanumber<0 \else\global\equanumber=0\fi
   \sectionnumber=0 \makel@bel}
\def\makel@bel{\xdef\chapterlabel{%
\the\chapterstyle{\the\chapternumber}.}}
\def\sectionlabel{\number\sectionnumber \quad }
\def\alphabetic#1{\count255='140 \advance\count255 by #1\char\count255}
\def\Alphabetic#1{\count255='100 \advance\count255 by #1\char\count255}
\def\Roman#1{\uppercase\expandafter{\romannumeral #1}}
\def\roman#1{\romannumeral #1}
\def\Number#1{\number #1}
\def\unnumberedchapters{\let\makel@bel=\relax \let\chapterlabel=\relax
\let\sectionlabel=\relax \equanumber=-1 }
\def\titlestyle#1{\par\begingroup \interlinepenalty=9999
     \leftskip=0.02\hsize plus 0.23\hsize minus 0.02\hsize
     \rightskip=\leftskip \parfillskip=0pt
     \hyphenpenalty=9000 \exhyphenpenalty=9000
     \tolerance=9999 \pretolerance=9000
     \spaceskip=0.333em \xspaceskip=0.5em
     \iftwelv@\twelvepoint\fourteenrm\else\twelvepoint\fi
   \noindent #1\par\endgroup }
\def\spacecheck#1{\dimen@=\pagegoal\advance\dimen@ by -\pagetotal
   \ifdim\dimen@<#1 \ifdim\dimen@>0pt \vfil\break \fi\fi}
\def\chapter#1{\par \penalty-300 \vskip\chapterskip
   \spacecheck\chapterminspace
   \chapterreset \titlestyle{\chapterlabel \ #1}
   \nobreak\vskip\headskip \penalty 30000
   \wlog{\string\chapter\ \chapterlabel} }

\def\section#1{\par \ifnum\the\lastpenalty=30000\else
   \penalty-200\vskip\sectionskip \spacecheck\sectionminspace\fi
   \wlog{\string\section\ \chapterlabel \the\sectionnumber}
   \global\advance\sectionnumber by 1  \noindent
   {\caps\enspace\chapterlabel \sectionlabel #1}\par
   \nobreak\vskip\headskip \penalty 30000 }
\def\subsection#1{\par
   \ifnum\the\lastpenalty=30000\else \penalty-100\smallskip \fi
   \noindent\undertext{#1}\enspace \vadjust{\penalty5000}}

\def\undertext#1{\vtop{\hbox{#1}\kern 1pt \hrule}}
\def\APPENDIX#1#2{\par\penalty-300\vskip\chapterskip
   \spacecheck\chapterminspace \chapterreset \xdef\chapterlabel{#1}
   \titlestyle{APPENDIX #2} \nobreak\vskip\headskip \penalty 30000
   \wlog{\string\Appendix\ \chapterlabel} }
\def\Appendix#1{\APPENDIX{#1}{#1}}
\def\appendix{\APPENDIX{A}{}}
%
%
%

\newif\ifdraftmode
\draftmodefalse

\def\eqname#1{\relax \ifnum\equanumber<0
     \xdef#1{{(\number-\equanumber)}}\global\advance\equanumber by -1
    \else \global\advance\equanumber by 1
      \xdef#1{{(\chapterlabel \number\equanumber)}} \fi}
\def\eqn#1{\eqno\eqname{#1}#1\ifdraftmode\rlap{\sevenrm\ \string#1}\fi}


%
\def\eqinsert#1{\noalign{\dimen@=\prevdepth \nointerlineskip
   \setbox0=\hbox to\displaywidth{\hfil #1}
   \vbox to 0pt{\vss\hbox{$\!\box0\!$}\kern-0.5\baselineskip}
   \prevdepth=\dimen@}}
%

%

%

%
%
\def\GENITEM#1;#2{\par \hangafter=0 \hangindent=#1
    \Textindent{$ #2 $}\ignorespaces}
\outer\def\newitem#1=#2;{\gdef#1{\GENITEM #2;}}
\newdimen\itemsize                \itemsize=30pt
\newitem\item=1\itemsize;
\newitem\sitem=1.75\itemsize;     
\newitem\ssitem=2.5\itemsize;     
\outer\def\newlist#1=#2&#3&#4;{\toks0={#2}\toks1={#3}%
   \count255=\escapechar \escapechar=-1
   \alloc@0\list\countdef\insc@unt\listcount     \listcount=0
   \edef#1{\par
      \countdef\listcount=\the\allocationnumber
      \advance\listcount by 1
      \hangafter=0 \hangindent=#4
      \Textindent{\the\toks0{\listcount}\the\toks1}}
   \expandafter\expandafter\expandafter
    \edef\c@t#1{begin}{\par
      \countdef\listcount=\the\allocationnumber \listcount=1
      \hangafter=0 \hangindent=#4
      \Textindent{\the\toks0{\listcount}\the\toks1}}
   \expandafter\expandafter\expandafter
    \edef\c@t#1{con}{\par \hangafter=0 \hangindent=#4 \noindent}
   \escapechar=\count255}
\def\c@t#1#2{\csname\string#1#2\endcsname}
\newlist\point=\Number&.&1.0\itemsize;
\newlist\subpoint=(\alphabetic&)&1.75\itemsize;
\newlist\subsubpoint=(\roman&)&2.5\itemsize;
\let\spoint=\subpoint

%
%
%
\newcount\referencecount     \referencecount=0
\newif\ifreferenceopen       \newwrite\referencewrite
\newtoks\rw@toks
\def\NPrefmark#1{\attach{\scriptscriptstyle [ #1 ] }}
\let\PRrefmark=\attach
\def\refmark#1{\relax\ifPhysRev\PRrefmark{#1}\else\NPrefmark{#1}\fi}
\def\refend{\refmark{\number\referencecount}}
\newcount\lastrefsbegincount \lastrefsbegincount=0
\def\refsend{\refmark{\count255=\referencecount
   \advance\count255 by-\lastrefsbegincount
   \ifcase\count255 \number\referencecount
   \or \number\lastrefsbegincount,\number\referencecount
   \else \number\lastrefsbegincount-\number\referencecount \fi}}
\def\refch@ck{\chardef\rw@write=\referencewrite
   \ifreferenceopen \else \referenceopentrue
   \immediate\openout\referencewrite=referenc.tex \fi}
%
{\catcode`\^^M=\active 
  \gdef\obeyendofline{\catcode`\^^M\active \let^^M\ }}%
%
{\catcode`\^^M=\active 
  \gdef\ignoreendofline{\catcode`\^^M=5}}
{\obeyendofline\gdef\rw@start#1{\def\t@st{#1} \ifx\t@st\blankend%
\endgroup \@sf \relax \else \ifx\t@st\bl@nkend \endgroup \@sf \relax%
\else \rw@begin#1
\backtotext
\fi \fi } }
{\obeyendofline\gdef\rw@begin#1
{\def\n@xt{#1}\rw@toks={#1}\relax%
\rw@next}}
\def\blankend{}
{\obeylines\gdef\bl@nkend{
}}
\newif\iffirstrefline  \firstreflinetrue
\def\rwr@teswitch{\ifx\n@xt\blankend \let\n@xt=\rw@begin %
 \else\iffirstrefline \global\firstreflinefalse%
\immediate\write\rw@write{\noexpand\obeyendofline \the\rw@toks}%
\let\n@xt=\rw@begin%
      \else\ifx\n@xt\rw@@d \def\n@xt{\immediate\write\rw@write{%
        \noexpand\ignoreendofline}\endgroup \@sf}%
             \else \immediate\write\rw@write{\the\rw@toks}%
             \let\n@xt=\rw@begin\fi\fi \fi}
\def\rw@next{\rwr@teswitch\n@xt}
\def\rw@@d{\backtotext} \let\rw@end=\relax
\let\backtotext=\relax

\newdimen\refindent     \refindent=30pt
\def\refitem#1{\par \hangafter=0 \hangindent=\refindent \Textindent{#1}}
\def\REFNUM#1{\space@ver{}\refch@ck \firstreflinetrue%
 \global\advance\referencecount by 1 \xdef#1{\the\referencecount}}
\def\refnum#1{\space@ver{}\refch@ck \firstreflinetrue%
 \global\advance\referencecount by 1 \xdef#1{\the\referencecount}\refend}

\def\REF#1{\REFNUM#1%
 \immediate\write\referencewrite{%
            \noexpand\refitem{\ifdraftmode{\sevenrm 
               \noexpand\string\string#1\ }\fi#1.}}%
      \begingroup\obeyendofline\rw@start}
\def\ref{\refnum\?%
 \immediate\write\referencewrite{\noexpand\refitem{\?.}}%
\begingroup\obeyendofline\rw@start}
\def\Ref#1{\refnum#1%
 \immediate\write\referencewrite{\noexpand\refitem{#1.}}%
\begingroup\obeyendofline\rw@start}
\def\REFS#1{\REFNUM#1\global\lastrefsbegincount=\referencecount
\immediate\write\referencewrite{\noexpand\refitem{#1.}}%
\begingroup\obeyendofline\rw@start}
%

%
%

%
\newcount\figurecount     \figurecount=0
\newif\iffigureopen       \newwrite\figurewrite
\def\figch@ck{\chardef\rw@write=\figurewrite \iffigureopen\else
   \immediate\openout\figurewrite=figures.aux
   \figureopentrue\fi}
\def\FIGNUM#1{\space@ver{}\figch@ck \firstreflinetrue%
 \global\advance\figurecount by 1 \xdef#1{\the\figurecount}}
\def\FIG#1{\FIGNUM#1
   \immediate\write\figurewrite{\noexpand\refitem{#1.}}%
   \begingroup\obeyendofline\rw@start}
\def\fig{\FIGNUM\? fig.~\?%
\immediate\write\figurewrite{\noexpand\refitem{\?.}}%
\begingroup\obeyendofline\rw@start}
\def\figure{\FIGNUM\? figure~\?
   \immediate\write\figurewrite{\noexpand\refitem{\?.}}%
   \begingroup\obeyendofline\rw@start}
\def\Fig{\FIGNUM\? Fig.~\?%
\immediate\write\figurewrite{\noexpand\refitem{\?.}}%
\begingroup\obeyendofline\rw@start}
\def\Figure{\FIGNUM\? Figure~\?%
\immediate\write\figurewrite{\noexpand\refitem{\?.}}%
\begingroup\obeyendofline\rw@start}
\newcount\tablecount     \tablecount=0
\newif\iftableopen       \newwrite\tablewrite
\def\tabch@ck{\chardef\rw@write=\tablewrite \iftableopen\else
   \immediate\openout\tablewrite=tables.aux
   \tableopentrue\fi}
\def\TABNUM#1{\space@ver{}\tabch@ck \firstreflinetrue%
 \global\advance\tablecount by 1 \xdef#1{\the\tablecount}}
\def\TABLE#1{\TABNUM#1
   \immediate\write\tablewrite{\noexpand\refitem{#1.}}%
   \begingroup\obeyendofline\rw@start}
\def\Table{\TABNUM\? Table~\?%
\immediate\write\tablewrite{\noexpand\refitem{\?.}}%
\begingroup\obeyendofline\rw@start}
\newif\ifsymbolopen       \newwrite\symbolwrite
\def\symch@ck{\ifsymbolopen\else
   \immediate\openout\symbolwrite=symbols.aux
   \symbolopentrue\fi}
\def\symdef#1#2{\def#1{#2}%
      \symch@ck%
      \immediate\write\symbolwrite{$$ \hbox{\noexpand\string\string#1} 
                \noexpand\qquad\noexpand\longrightarrow\noexpand\qquad 
                           \string#1 $$}}
%
%

%
%
%
\def\masterreset{\global\pagenumber=1 \global\chapternumber=0
   \global\equanumber=0 \global\sectionnumber=0
   \global\referencecount=0 \global\figurecount=0 \global\tablecount=0 }
\def\FRONTPAGE{\ifvoid255\else\vfill\penalty-2000\fi
      \masterreset\global\frontpagetrue
      \global\lastf@@t=0 \global\footsymbolcount=0}

\def\paperstyle{\letterstylefalse\normalspace\papersize}
\def\letterstyle{\letterstyletrue\singlespace\lettersize}
\def\papersize{\hsize=35pc\vsize=50pc\hoffset=1pc\voffset=6pc
               \skip\footins=\bigskipamount}
\def\lettersize{\hsize=6.5in\vsize=8.5in\hoffset=0in\voffset=1in
   \skip\footins=\smallskipamount \multiply\skip\footins by 3 }
\paperstyle   
%
%
\def\MEMO{\letterstyle\FRONTPAGE \letterfrontheadline={\hfil}
    \line{\quad\fourteenrm NTC MEMORANDUM\hfil\twelverm\the\date\quad}
    \medskip \memod@f}

\def\memit@m#1{\smallskip \hangafter=0 \hangindent=1in
      \Textindent{\caps #1}}
\def\memod@f{\xdef\to{\memit@m{To:}}\xdef\from{\memit@m{From:}}%
     \xdef\topic{\memit@m{Topic:}}\xdef\subject{\memit@m{Subject:}}%
     \xdef\rule{\bigskip\hrule height 1pt\bigskip}}
\memod@f
\newskip\lettertopfil
\lettertopfil = 0pt plus 1.5in minus 0pt
\newskip\letterbottomfil
\letterbottomfil = 0pt plus 2.3in minus 0pt
\newskip\spskip \setbox0\hbox{\ } \spskip=-1\wd0
\def\addressee#1{\medskip\rightline{\the\date\hskip 30pt} \bigskip
   \vskip\lettertopfil
   \ialign to\hsize{\strut ##\hfil\tabskip 0pt plus \hsize \cr #1\crcr}
   \medskip\noindent\hskip\spskip}
\newskip\signatureskip       \signatureskip=40pt
\def\signed#1{\par \penalty 9000 \bigskip \dt@pfalse
  \everycr={\noalign{\ifdt@p\vskip\signatureskip\global\dt@pfalse\fi}}
  \setbox0=\vbox{\singlespace \halign{\tabskip 0pt \strut ##\hfil\cr
   \noalign{\global\dt@ptrue}#1\crcr}}
  \line{\hskip 0.5\hsize minus 0.5\hsize \box0\hfil} \medskip }

\def\endletter{\ifnum\pagenumber=1 \vskip\letterbottomfil\supereject
\else \vfil\supereject \fi}
\newbox\letterb@x
\def\lettertext{\par\unvcopy\letterb@x\par}
\def\multiletter{\setbox\letterb@x=\vbox\bgroup
      \everypar{\vrule height 1\baselineskip depth 0pt width 0pt }
      \singlespace \topskip=\baselineskip }
\def\letterend{\par\egroup}
%
%
%
\newskip\frontpageskip
\newtoks\pubtype
\newtoks\Pubnum
\newtoks\pubnum
\newif\ifp@bblock  \p@bblocktrue
\def\PH@SR@V{\doubl@true \baselineskip=24.1pt plus 0.2pt minus 0.1pt
             \parskip= 3pt plus 2pt minus 1pt }
\def\PHYSREV{\paperstyle\PhysRevtrue\PH@SR@V}
\def\titlepage{\FRONTPAGE\paperstyle\ifPhysRev\PH@SR@V\fi
   \ifp@bblock\p@bblock\fi}
\def\nopubblock{\p@bblockfalse}
\def\endpage{\vfil\break}
\frontpageskip=1\medskipamount plus .5fil
\pubtype={\tensl Preliminary Version}
\Pubnum={$\caps SLAC - PUB - \the\pubnum $}
\pubnum={0000}
\def\p@bblock{\begingroup \tabskip=\hsize minus \hsize
   \baselineskip=1.5\ht\strutbox \topspace-2\baselineskip
   \halign to\hsize{\strut ##\hfil\tabskip=0pt\crcr
   \the\Pubnum\cr \the\date\cr \the\pubtype\cr}\endgroup}
\def\title#1{\vskip\frontpageskip \titlestyle{#1} \vskip\headskip }
\def\author#1{\vskip\frontpageskip\titlestyle{\twelvecp #1}\nobreak}

\def\address#1{\par\kern 5pt\titlestyle{\twelvepoint\it #1}}
\def\andaddress{\par\kern 5pt \centerline{\sl and} \address}

\def\abstract{\vskip\frontpageskip\centerline{\fourteenrm ABSTRACT}
              \vskip\headskip }

%
%
%

\def\\{\relax\ifmmode\backslash\else$\backslash$\fi}
\def\globaleqnumbers{\relax\if\equanumber<0\else\global\equanumber=-1\fi}

\def\journal#1&#2(#3){\unskip, \sl #1~\bf #2 \rm (19#3) }

\def\topspace{\hrule height 0pt depth 0pt \vskip}

\let\int=\intop         
\def\prop{\mathrel{{\mathchoice{\pr@p\scriptstyle}{\pr@p\scriptstyle}{
                \pr@p\scriptscriptstyle}{\pr@p\scriptscriptstyle} }}}
\def\pr@p#1{\setbox0=\hbox{$\cal #1 \char'103$}
   \hbox{$\cal #1 \char'117$\kern-.4\wd0\box0}}
\def\lsim{\mathrel{\mathpalette\@versim<}}
\def\gsim{\mathrel{\mathpalette\@versim>}}
\def\@versim#1#2{\lower0.5ex\vbox{\baselineskip\z@skip\lineskip-.1ex
  \lineskiplimit\z@\ialign{$\m@th#1\hfil##\hfil$\crcr#2\crcr\sim\crcr}}}
%
%
%
\let\sec@nt=\sec
\def\sec{\relax\ifmmode\let\n@xt=\sec@nt\else\let\n@xt\section\fi\n@xt}
\def\obsolete#1{\message{Macro \string #1 is obsolete.}}
\def\firstsec#1{\obsolete\firstsec \section{#1}}
\def\firstsubsec#1{\obsolete\firstsubsec \subsection{#1}}
\def\thispage#1{\obsolete\thispage \global\pagenumber=#1\frontpagefalse}
\def\thischapter#1{\obsolete\thischapter \global\chapternumber=#1}
\def\nextequation#1{\obsolete\nextequation \global\equanumber=#1
   \ifnum\the\equanumber>0 \global\advance\equanumber by 1 \fi}
\def\BOXITEM{\afterassigment\B@XITEM\setbox0=}
\def\B@XITEM{\par\hangindent\wd0 \noindent\box0 }
%

%
%
%
%
%
\lock
\message{   }
%
%
%












\newbox\figbox
\newdimen\zero  \zero=0pt
\newdimen\figmove
\newdimen\figwidth
\newdimen\figheight
\newdimen\textwidth
\newtoks\figtoks
\newcount\figcounta
\newcount\figcountb
\newcount\figlines
\def\figreset{\global\figcounta=-1 \global\figcountb=-1
\global\figmove=\baselineskip
\global\figlines=1 \global\figtoks={ } }
\def\picture#1by#2:#3{\global\setbox\figbox=\vbox{\vskip #1
\hbox{\vbox{\hsize=#2 \noindent #3}}}
\global\setbox\figbox=\vbox{\kern 10pt
\hbox{\kern 10pt \box\figbox \kern 10pt }\kern 10pt}
\global\figwidth=1\wd\figbox
\global\figheight=1\ht\figbox
\global\textwidth=\hsize
\global\advance\textwidth by - \figwidth }
\def\figtoksappend{\edef\temp##1{\global\figtoks=%
{\the\figtoks ##1}}\temp}
\def\figparmsa#1{\loop \global\advance\figcounta by 1
\ifnum \figcounta < #1
\figtoksappend{ 0pt \the\hsize }
\global\advance\figlines by 1
\repeat }
\def\figparmsb#1{\loop \global\advance\figcountb by 1
\ifnum \figcountb < #1
\figtoksappend{ \the\figwidth \the\textwidth}
\global\advance\figlines by 1
\repeat }
\def\figtext#1:#2:#3{\figreset%
\figparmsa{#1}%
\figparmsb{#2}%
\multiply\figmove by #1%
\global\setbox\figbox=\vbox to 0pt{\vskip \figmove  \hbox{\box\figbox}
\vss }
\parshape=\the\figlines\the\figtoks\the\zero\the\hsize
\noindent
\rlap{\box\figbox} #3}
\def\Buildrel#1\under#2{\mathrel{\mathop{#2}\limits_{#1}}}
\def\llongrarrow{\hbox to 40pt{\rightarrowfill}}



\def\boxit#1{\vbox{\hrule\hbox{\vrule\kern3pt
\vbox{\kern3pt#1\kern3pt}\kern3pt\vrule}\hrule}}
\newdimen\str
\def\fboxit#1#2{\vbox{\hrule height #1 \hbox{\vrule width #1
\kern3pt \vbox{\kern3pt#2\kern3pt}\kern3pt \vrule width #1 }
\hrule height #1 }}
\def\tran#1#2{\transpoint \hfuzz 5pt \gdef\label{#1}
\vbox to \the\vsize{\hsize \the\hsize #2} \par \eject }
\newdimen\baseskip
\newdimen\lskip
\lskip=\lineskip
\newdimen\transize
\newdimen\tall
\def\transpoint{\gdef\rm{\fam0\eighteenrm}
\font\twentyfourit = cmti10 scaled \magstep5
\font\twentyfourrm = cmr10 scaled \magstep5
\font\twentyfourbf = cmbx10 scaled \magstep5
\font\twentyeightsy = cmsy10 scaled \magstep5
\font\eighteenrm = cmr10 scaled \magstep3
\font\eighteenb = cmbx10 scaled \magstep3
\font\eighteeni = cmmi10 scaled \magstep3
\font\eighteenit = cmti10 scaled \magstep3
\font\eighteensl = cmsl10 scaled \magstep3
\font\eighteensy = cmsy10 scaled \magstep3
\font\eighteencaps = cmr10 scaled \magstep3
\font\eighteenmathex = cmex10 scaled \magstep3
\font\fourteenrm=cmr10 scaled \magstep2
\font\fourteeni=cmmi10 scaled \magstep2
\font\fourteenit = cmti10 scaled \magstep2
\font\fourteensy=cmsy10 scaled \magstep2
\font\fourteenmathex = cmex10 scaled \magstep2
\parindent 20pt
\global\hsize = 7.0in
\global\vsize = 8.9in
\dimen\transize = \the\hsize
\dimen\tall = \the\vsize
\def\sy{\eighteensy }
\def\sl{\eighteens }
\def\bf{\eighteenb }
\def\it{\eighteenit }
\def\caps{\eighteencaps }
\textfont0=\eighteenrm \scriptfont0=\fourteenrm
\scriptscriptfont0=\twelverm
\textfont1=\eighteeni \scriptfont1=\fourteeni \scriptscriptfont1=\twelvei
\textfont2=\eighteensy \scriptfont2=\fourteensy
\scriptscriptfont2=\twelvesy
\textfont3=\eighteenmathex \scriptfont3=\eighteenmathex
\scriptscriptfont3=\eighteenmathex
\global\baselineskip 35pt
\global\lineskip 15pt
\global\parskip 5pt  plus 1pt minus 1pt
\global\abovedisplayskip  3pt plus 10pt minus 10pt
\global\belowdisplayskip 3pt plus 10pt minus 10pt
\def\rtitle##1{\centerline{\undertext{\twentyfourrm ##1}}}
\def\ititle##1{\centerline{\undertext{\twentyfourit ##1}}}
\def\ctitle##1{\centerline{\undertext{\caps ##1}}}
\def\vstrut{\hbox{\vrule width 0pt height .35in depth .15in }}
\def\cline##1{\centerline{\vstrut ##1}}
\output{\shipout\vbox{\vskip .5in
\pagecontents \vfill
\hbox to \the\hsize{\hfill{\tenbf \label} } }
\global\advance\count0 by 1 }
\rm }


%
%
%

%

%

%

%
{\obeyspaces\global\let =\ }
%
%
%
\widowpenalty 1000
\thickmuskip 4mu plus 4mu

\def\bspac{\noalign{\vskip 4pt}}

\unlock
%
\Pubnum={${\twelverm IU/NTC}\  \the\pubnum $}
\pubnum={0000}
\def\p@nnlock{\begingroup \tabskip=\hsize minus \hsize
   \baselineskip=1.5\ht\strutbox \topspace-2\baselineskip
   \noindent\strut\the\Pubnum \hfill \the\date   \endgroup}
\def\titlepage{\FRONTPAGE\paperstyle\p@nnlock}
\def\displaylines#1{\displ@y
  \halign{\hbox to\displaywidth{$\hfil\displaystyle##\hfil$}\crcr
    #1\crcr}}
\def\addressee#1{\null
   \bigskip\medskip\rightline{\the\date\hskip 30pt}
   \vskip\lettertopfil
   \ialign to\hsize{\strut ##\hfil\tabskip 0pt plus \hsize \cr #1\crcr}
   \medskip\vskip 3pt\noindent}
\def\tmsaddressee#1#2{
   \vskip\lettertopfil
  \setbox0=\vbox{\singlespace \halign{\tabskip 0pt \strut ##\hfil\cr
   \noalign{\global\dt@ptrue}#1\crcr}}
  \line{\hskip 0.7\hsize minus 0.7\hsize \box0\hfil}
   \bigskip
   \vskip .2in
   \ialign to\hsize{\strut ##\hfil\tabskip 0pt plus \hsize \cr #2\crcr}
   \medskip\vskip 3pt\noindent}
\def\makeheadline{\vbox to 0pt{ \skip@=\topskip
      \advance\skip@ by -12pt \advance\skip@ by -2\normalbaselineskip
      \vskip\skip@  \vss
      }\nointerlineskip}
\def\signed#1{\par \penalty 9000 \bigskip \vskip .06in\dt@pfalse
  \everycr={\noalign{\ifdt@p\vskip\signatureskip\global\dt@pfalse\fi}}
  \setbox0=\vbox{\singlespace \halign{\tabskip 0pt \strut ##\hfil\cr
   \noalign{\global\dt@ptrue}#1\crcr}}
  \line{\hskip 0.5\hsize minus 0.5\hsize \box0\hfil} \medskip }
\def\lettersize{\hsize=6.25in\vsize=8.5in\hoffset=0in\voffset=1in
   \skip\footins=\smallskipamount \multiply\skip\footins by 3 }
%
%
%
%
%
\outer\def\newnewlist#1=#2&#3&#4&#5;{\toks0={#2}\toks1={#3}%
   \dimen1=\hsize  \advance\dimen1 by -#4
   \dimen2=\hsize  \advance\dimen2 by -#5
   \count255=\escapechar \escapechar=-1
   \alloc@0\list\countdef\insc@unt\listcount     \listcount=0
   \edef#1{\par
      \countdef\listcount=\the\allocationnumber
      \advance\listcount by 1
      \parshape=2 #4 \dimen1 #5 \dimen2
      \Textindent{\the\toks0{\listcount}\the\toks1}}
   \expandafter\expandafter\expandafter
    \edef\c@t#1{begin}{\par
      \countdef\listcount=\the\allocationnumber \listcount=1
      \parshape=2 #4 \dimen1 #5 \dimen2
      \Textindent{\the\toks0{\listcount}\the\toks1}}
   \expandafter\expandafter\expandafter
    \edef\c@t#1{con}{\par \parshape=2 #4 \dimen1 #5 \dimen2 \noindent}
   \escapechar=\count255}
\def\c@t#1#2{\csname\string#1#2\endcsname}
%
%
%
%
%
%
%
\def\noparGENITEM#1;{\hangafter=0 \hangindent=#1
    \ignorespaces\noindent}
\outer\def\noparnewitem#1=#2;{\gdef#1{\noparGENITEM #2;}}
\noparnewitem\spoint=1.5\itemsize;
%
%
%
\def\MEMO{\letterstyle\FRONTPAGE \letterfrontheadline={\hfil}
      \hoffset=1in \voffset=1.21in
    \line{\hskip .8in  \special{overlay ntcmemo.dat}
          \quad\fourteenrm NTC MEMORANDUM\hfil\twelverm\the\date\quad}
    \medskip\medskip \memod@f}

\def\memit@m#1{\smallskip \hangafter=0 \hangindent=1in
      \Textindent{\caps #1}}
\def\memod@f{\xdef\to{\memit@m{To:}}\xdef\from{\memit@m{From:}}%
     \xdef\topic{\memit@m{Topic:}}\xdef\subject{\memit@m{Subject:}}%
     \xdef\rule{\bigskip\hrule height 1pt\bigskip}}
\memod@f
\lock
%

%
%
%
%
%
%
%

\let\ssize=\scriptstyle
\let\sssize\scriptscriptstyle
%
%
%
%
%
\def\tildesymbol{\mathchar"0218 }
\def\undervec#1{\setbox0\hbox{$\tildesymbol$}\setbox1\hbox{$#1$}#1%
                \dimen0=\wd0\advance\dimen0by\wd1\divide\dimen0by2
                 \kern-\dimen0\lower1.3ex\hbox{$\tildesymbol$}}
\def\sundervec#1{\setbox0\hbox{$\ssize\tildesymbol$}
                 \setbox1\hbox{$\ssize #1$}#1%
                \dimen0=\wd0\advance\dimen0by\wd1\divide\dimen0by2
                 \kern-\dimen0\lower.85ex\hbox{$\ssize\tildesymbol$}}
\def\ssundervec#1{\setbox0\hbox{$\sssize\tildesymbol$}
                 \setbox1\hbox{$\sssize #1$}#1%
                \dimen0=\wd0\advance\dimen0by\wd1\divide\dimen0by2
                 \kern-\dimen0\lower.6ex\hbox{$\sssize\tildesymbol$}}
%

%


%

%

%

%

%

%

%

%
%
%
%
%

%

%

%

%

%

%
%
%
%
%
%

%

%

%

%

%
%
%
%
%
\def\intback{\kern-.1em}

%
%
%
%
%
\def\threej#1#2#3#4#5#6{\left(\matrix{#1 & #2 & #3 \cr
                                      #4 & #5 & #6 \cr}\right)}
\def\sixj#1#2#3#4#5#6{\left\{\matrix{#1 & #2 & #3 \cr
                                      #4 & #5 & #6 \cr}\right\}}

\def\bspac{\noalign{\vskip 4pt}}

%
%
%
%
%

%
%
%
%
%

%
%
%
%
%

%

%
%

%

%

%

%

%

%

%

%

%
%
%
%
%

%
%
%
%

%
%
%
\def\papersize{\hsize=6.25in\vsize=8.60in\hoffset=0.0in\voffset=0.1in
                \skip\footins=\bigskipamount}
\paperstyle
\doublespace
\pretolerance=500
\tolerance=500
\widowpenalty 5000
\thinmuskip=3mu
\medmuskip=4mu plus 2mu minus 4mu
\thickmuskip=5mu plus 5mu
\def\NPrefmark#1{\attach{\ssize #1{}}}
%
%
%
%
\def\NPrefmark#1{\attach{\scriptstyle  #1 }}

\def\Ll{{\lambda}}
\def\Llo{{\lambda_1}}
\def\Llt{{\lambda_2}}
\def\Llth{{\lambda_3}}

\PHYSREV
%
%

\REF\WATSON{G.~N.~Watson, {\it A Treatise on the Theory of Bessel
   Functions}, (Cambridge Univ. Press, 1944).}  

\REF\TOBOC{T.~Sawaguri and W.~Tobocman, J.~Math.~Phys. {\bf 8}, 
2223 (1967).}

\REF\JACKSON{A.~D.~Jackson and L.~Maximon, SIAM J.~Math.~Anal.~{\bf 3},
   446 (1972).}

\REF\TAFFAR{R.~Anni and L.~Taffara, Nuovo Cimento A {\bf 22}, 11 (1974).}

\REF\ELBAZM{E.~Elbaz, J.~Meyer and R.~Nahabetian, Lett.~Nuovo Cimento
{\bf 10}, 418 (1974).}

\REF\ELBAZ{E.~Elbaz, Riv.~Nuovo Cimento {\bf 5}, 561 (1975).}

\REF\GERV{A.~Gervois and H.~Navelet, J.~Math.~Phys.~{\bf 26}, 633 (1985).}

\REF\GERVA{A.~Gervois and H.~Navelet, SIAM J.~Math.~Anal.~{\bf 20}, 1006 (1989)}

\REF\WHITE{G.~D.~White, K.~T.~R.~Davies and P.~J.~Siemens, Ann.~Phys. (NY) {\bf
187}, 198 (1988).}

\REF\DAVIES{K.~T.~R.~Davies, M.~R.~Strayer and G.~D.~White, J.~Phys.~G {\bf 14},
961 (1988).}

\REF\DAVIEST{K.~T.~R.~Davies, J.~Phys.~G {\bf 14}, 973 (1988).}

\REF\MEHR{R.~Mehrem, J.~T.~Londergan and M.~H.~Macfarlane, J.~Phys.~A {\bf 24},
1435 (1991).}

\REF\MEHREM{ R.~Mehrem, {\it Ph.D.~Thesis}, (Indiana University, 1992)
(unpublished).}

\REF\MEHRE{R.~Mehrem, J.~T.~Londergan and G.~E.~Walker, Phys.~Rev.~C {\bf 48},
1192 (1993).}

\REF\WHE{C.T.~Whelan, J.~Phys.~B {\bf 26}, 823 (1993).}

\REF\CHEN{J.~Chen and J.~Su, Phys.~Rev.~D {\bf 69}, 076003 (2004).}

\REF\JORISA{J.~Van Deun and R.~Cools, ACM Trans.~Math.~Softw.~{\bf 32}, 580 (2006).}

\REF\JORIS{J.~Van Deun and R.~Cools, Comput.~Phys.~Comm.~{\bf 178}, 578 (2008).}

\REF\HOH{A.~Hohenegger, Phys.~Rev.~D {\bf 79}, 063502 (2009), arXiv:0806.3098.}

\REF\HOHDIS{A.~Hohenegger, {\it Ph.D.~Thesis}, (University of Heidelberg,
2010).}

\REF\GEB{B.~Gebremariam, T.~Duguet and S.~K.~Bogner, Comput.~Phys.~Comm. {\bf
181}, 1136 (2010).}

\REF\MEH{R.~Mehrem, arXiv:math-ph/09090494.}

\REF\BRNKSA{\rm D.~M.~Brink and G.~R.~Satchler, {\it Angular Momentum} (Oxford
University Press, London, 1962).}

\REF\EDMNDS{\rm A.~R.~Edmonds, {\it Angular Momentum in Quantum Mechanics}
(Princeton University Press, Princeton, 1957).}

\REF\GRADST{\rm I.~S.~Gradshteyn and I.~M.~Rhyzik: {\it Table of Integrals, 
Series and Products} (Academic Press, New York, Fifth Edition, 1994).} 

\REF\FIN{ S.~Fineschi, E.~Landi and Degl'Innocenti, 
J.~Math.~Phys. {\bf 31}, 1124 (1990).} 

\REF\VIR{\rm N.~Virchenko and I.~Fedotova, {\it Generalized Associated Legendre
Functions And Their Applications} (World Scientific, Singapore, 2001).}

\REF\INPREP{A.~Hohenegger and R.~Mehrem, {\it in preperation}.}

\title{\bf A GENERALISATION FOR THE INFINITE INTEGRAL OVER THREE SPHERICAL
BESSEL FUNCTIONS}
\author{R.\ Mehrem$^{\bf a}$} 
\address{\it Associate Lecturer \break
The Open University in the North West \break
351 Altrincham Road \break
Sharston, Manchester M22 4UN \break
United Kingdom}

\author{A.\ HOHENEGGER$^{\bf b}$}
\address{\it Max-Planck-Institut f\"ur Kernphysik \break
Saupfercheckweg 1 \break
69117 Heidelberg \break
Germany}
\footnote{}{\bf a Email: rami.mehrem@btopenworld.com.} 
\footnote{}{\bf b Email: Andreas.Hohenegger@mpi-hd.mpg.de.} 
\endpage
\abstract{A new formula is derived that generalises an earlier result for the
infinite integral over three spherical Bessel functions. The analytical result
involves a finite sum over associated Legendre functions, $P_l^m(x)$ of degree
$l$ and order $m$. The sum allows for values of $|\,m|$ that are greater than
$l$. A generalisation for the associated Legendre functions to allow for any
rational $m$ for a specific $l$ is also shown.}
 
\endpage
%
%
\chapter{Introduction}
The calculations of infinite integrals that involve a product of spherical
Bessel functions have been the focus of many papers, amongst which are
references [\WATSON -\MEH]. In particular, reference [\MEHR] showed a derivation
for an analytical evaluation of
an infinite integral over three spherical Bessel functions, given by the form 

$$I(\Llo,\,\Llt,\,\Llth\,;\,k_1\,k_2\,k_3)\,\equiv \,\int_0^\infty \,
r^2\,j_{\Llo}(k_1r)\,j_{\Llt}(k_2r)\,j_{\Llth}(k_3r)\,dr. 
\eqn\ONEPONE$$ 
This type of integral has received considerable attention due to its use in
nuclear scattering theory 
[\WHITE -\MEHRE]. In this paper, we extend this result to analytically evaluate
infinite integrals of the form 

$$I(\Ll\,;\,\Llo,\,\Llt,\,\Llth\,;\,k_1\,k_2\,k_3)\,\equiv \,
\int_0^\infty\,r^{2-\Ll}\,j_{\Llo}(k_1r)\,j_{\Llt}(k_2r)\,j_{\Llth+\Ll}(k_3r)\,
dr.\eqn\ONEPTWO$$ 
This type of integral has in the past been attempted both analytically [\GERVA] and numerically [\JORISA]. 
The authors believe that the evaluation carried out here is more compact, easier to work with and extendible. 
The new result, like the earlier one when $\lambda\,=\,0$, only works when $|\Llo - \Llt| \le \Llth
\le \Llo + \Llt$, i.e.~when the integer indices 
$\Llo$, $\Llt$ and $\Llth$ satisfy the triangular condition, and $\Llo + \Llt + \Llth$ is even. However, unlike the
earlier result where the triangular condition forces 
$k_1$, $k_2$ and $k_3$ to form the sides of a triangle, this generalised
integral can have non-zero values when the $k$'s do not form the sides of a
triangle. Our new result is only derived for the case the $k$'s do form the
sides of a triangle. We also show the general result, i.e.~for any values of
$k_1$, $k_2$ and $k_3$, when 
$\Ll\,=\,1$.
This new result involves finite sums over the associated Legendre function,
$P_l^m(x)$, of degree $l$ and order $m$, whereas the earlier result involved
sums over the Legendre polynomial, $P_l(x)$. The sums can involve values for
$|m|$ that are larger than $l$. An extension for the associated Legendre
functions which produces a formula that allows for such values is shown in appendix A. 
An application is shown to evaluate an integral involving four
spherical Bessel functions and is used in the angular integration of the
homogeneous and velocity isotropic Boltzmann equation. This integral, which is
highly oscillatory and needs special treatment when evaluated numerically, is
reduced to an integral over
an associated Legendre function and a Legendre polynomial combined with
algebraic factors which can easily be evaluated numerically. 
A comparison is shown in appendix B between our analytical result and the analytical evaluation carried out in 
reference [\GERVA].

\chapter{Generalising the Integral Over Three Spherical Bessel Functions}
 
An earlier result [\MEHR] showed that an infinite integral over three spherical
Bessel functions can be written as
$$\eqalign{&\threej{\Llo}{\Llt}{\Llth}
{0}{0}{0}\,I(\Llo,\,\Llt,\,\Llth\,;\,k_1\,k_2\,k_3)\,
=\,{\pi\beta(\Delta) \over 4 k_1 k_2 k_3} i^{\Llo+\Llt-\Llth}\cr\bspac
&\times(2\Llth+1)^{1/2}
\left({k_1 \over k_3}\right)^\Llth \,\sum_{{\cal L}=0}^\Llth\,{2\Llth \choose
2{\cal L}}^{1/2} 
\left({k_2 \over k_1}\right)^{\cal L} 
\sum_l\, (2l+1)\,
\threej{\Llo}{\Llth - {\cal L}}{l}{0}{0}{0}\cr\bspac
&\times\threej{\Llt}{{\cal L}}{l}{0}{0}{0}\,
\sixj{\Llo}{\Llt}{\Llth}{{\cal L}}{\Llth-{\cal L}}{l}\,
P_l(\Delta),\cr }\eqn\TWOPONE$$
where $\Delta \,=\,(k_1^2+k_2^2-k_3^2)/2k_1k_2$ and 
$\beta (\Delta)=\theta (1-\Delta)\theta (1+\Delta)$ with $\theta$ the Heaviside function 
in half-maximum convention. $P_l(x)$ is a Legendre
polynomial of order $l$, $\threej{\Llo}{\Llt}{\Llth}
{0}{0}{0}$ is a 3j symbol and $\sixj{\Llo}{\Llt}{\Llth}{{\cal L}}{\Llth-{\cal
L}}{l}$ is a 6j symbol which can be found in any standard angular momentum text
[\BRNKSA , \EDMNDS]. Note that the summand in $l$ vanishes unless
$|\Llo\,-\,(\Llth\,-\,{\cal L})| \,\le\,l\, \le \Llo\,+\,\Llth\,-\,{\cal L}$.
Now, multiply both sides by $k_3^{\Llth+2}$ and integrate over 
$k_3$ from 0 to $K$, where $K$ can have any positive value. The left hand side involves
the integral (see [\GRADST], eq.~5.52-1, page 661)
$$\int_0^K\,k_3^{\Llth+2}\,j_\Llth(k_3r)\, dk_3\,=\,{1\over
r}\,K^{\Llth+2}\,j_{\Llth +1}(Kr).\eqn\TWOPTWO$$
The right hand side involves the integral
$$J\,\equiv \,\int_0^K \beta(\Delta)\,k_3\,P_l(\Delta)\,dk_3.\eqn\TWOPTHREE$$
Now, $k_3\,dk_3 \,=\,-k_1k_2\,d\Delta$. So, when $k_3\,=\,0$ then 
$k_1=k_2$ and $\Delta\,=\,1$. Also, when $k_3\,=\,K$,
$\Delta\,=\Delta ^\prime$, where $\Delta^\prime\,=\,
(k_1^2+k_2^2-K^2)/2k_1k_2$. Hence
$$J\,=\,k_1k_2\,\int_{\Delta^\prime}^1\,\beta(\Delta)
P_l(\Delta)\,d\Delta .\eqn\TWOPFOUR$$
If $\Delta^\prime > 1$, i.e. $K<|k_2-k_1|$, then $J=0$. If $-1\,\le
\Delta^\prime\,\le \,1$, then 
$$J\,=\,k_1k_2\,\beta(\Delta^\prime)\,(1-{\Delta^\prime}^2)^{1/2}\,P_l^{-1}
(\Delta^\prime),\eqn\TWOPFIVE$$
using [15]
$$P_l^{-m}(x)\,=\,(1-x^2)^{-m/2}\,\int_x^1\,...\int_x^1\,
P_l(x)\,(dx)^m,\eqn\TWOPSIX$$
where $P_l^m(x)$ is the associated Legendre function of the first kind of degree
$l$ and order $m$. If $\Delta^\prime <-1$, then
$$J\,=\,k_1k_2\,\theta[K-(k_1+k_2)]\,\int_{-1}^1\,P_l(\Delta)\,
d\Delta\,=\,2k_1k_2\,\theta[K-(k_1+k_2)]\,\delta_{l,\,\,0}
.\eqn\TWOPSEVEN$$

Hence, the result is
$$\eqalign{&\threej{\Llo}{\Llt}{\Llth}
{0}{0}{0}\,I(1\,;\,\Llo,\,\Llt,\,\Llth\,;\,k_1\,k_2\,K)\,
=\,{\pi\,\beta(\Delta^\prime) \over 4 K^2}\, i^{\Llo+\Llt-\Llth}\cr\bspac &
\times(2\Llth+1)^{1/2}\,
\left({k_1 \over K}\right)^\Llth  (1-{\Delta^\prime}^2)^{1/2}\,
\sum_{{\cal L}=0}^\Llth\,{2\Llth \choose 2{\cal L}}^{1/2}\, 
\left({k_2 \over k_1}\right)^{\cal L} 
\cr\bspac & \times\sum_l\,(2l+1)\,
\threej{\Llo}{\Llth - {\cal L}}{l}{0}{0}{0} \,\threej{\Llt}{{\cal
L}}{l}{0}{0}{0}\,
\sixj{\Llo}{\Llt}{\Llth}{{\cal L}}{\Llth-{\cal L}}{l}\,
P_l^{-1}(\Delta^\prime)\cr\bspac &+\,(-1)^\Llth\,{\pi\over
2}\,{k_1^\Llo\,k_2^\Llt \over K^{\Llth+2}}
\,{\sqrt{2\Llth\,+\,1}\over(2\Llo\,+\,1)\,(2\Llt\,+\,1)}\,{2\Llth \choose
2\Llt}^{1/2}\, 
\theta[K-(k_1+k_2)]\,\delta_{\Llth,\,\,\Llo+\Llt}.\cr }
\eqn\TWOPEIGHT$$

The result $\TWOPEIGHT$ applies for any values of 
$k_1$, $k_2$ and $K$. The second term on the right hand side of $\TWOPEIGHT$ is also consistent with equation 6.578-4 of 
reference [\GRADST] when $\Llth\,=\,\Llo\,+\,\Llt$. This equation applies for $K\,> \,k_1\,+k_2$ and can be written in the form

$$\eqalign{&\int_0^\infty\,r^{\Llth-\Llo-\Llt+1}\,j_\Llo (k_1r)\,j_\Llt(k_2r)\,
j_{\Llth+1}(Kr)dr\,=\,2^{\Llth-\Llt-\Llo-2}\,\pi^{3/2}\cr\bspac 
&\times {k_1^\Llo\,k^\Llt \over K^{\Llth+2}}\,{\Gamma(\Llth+3/2)\over \Gamma(\Llt+3/2)\,\Gamma(\Llo+3/2)}~~.}\eqn\TWOPNINE$$

If we assume that $K$ satisfies the triangular condition, then the second term in $\TWOPEIGHT$ vanishes, and by repeated integration one arrives at (renaming $K$ back to
$k_3$)
$$\eqalign{&I(\Ll\,;\,\Llo,\,\Llt,\,\Llth\,;\,k_1\,k_2\,k_3)\,
=\,{\pi\beta(\Delta) \over 4 k_1 k_2 k_3} i^{\Llo+\Llt-\Llth}\,(2\Llth+1)^{1/2}
\left({k_1 \over k_3}\right)^\Llth  \cr\bspac & \times\left ({k_1k_2 \over
k_3}\right )^\Ll \, (1-\Delta^2)^{\Ll/2}\,
{\threej{\Llo}{\Llt}{\Llth}
{0}{0}{0}}^{-1}\,\sum_{{\cal L}=0}^\Llth\,{2\Llth \choose 2{\cal L}}^{1/2} 
\left({k_2 \over k_1}\right)^{\cal L} 
\cr\bspac & \times\sum_l\,(2l+1)\,
\threej{\Llo}{\Llth - {\cal L}}{l}{0}{0}{0} \,\threej{\Llt}{{\cal
L}}{l}{0}{0}{0}\,
\sixj{\Llo}{\Llt}{\Llth}{{\cal L}}{\Llth-{\cal L}}{l}\,
P_l^{-\Ll}(\Delta),\cr}\eqn\TWOPTEN $$
where $\Ll$, $\Llo$, $\Llt$ and $\Llth$ are all greater than or equal to $0$ to
ensure convergence of the integral. Equation $\TWOPTEN$ is a new result that
generalizes the infinite integral over three spherical Bessel functions. 
In general $\lambda$ can exceed the value for $l$. In this case the definition
of $P_l^m(x)$ needs to be extended for 
$|\,m| > l$ in a way compatible with eq.~$\TWOPSIX$. 
Appendix A discusses such a generalization for the associated Legendre
functions. Several explicit results for eq.~$\TWOPTEN$ are presented in appendix B.
For some small values of the indices these have been compared with a different existing 
result for the integral as will also be discussed there.

\chapter{Special Cases and Identities}
Equation $\TWOPTEN$ can be reduced to a known integral [\JACKSON] for the case
$\Llth=0$, i.e. $\Llo\,=\,\Llt\,\equiv\,\lambda^\prime$ with the result 
$$I(\Ll\,;\,\Ll^\prime,\,\Ll^\prime,\,0\,;\,k_1\,k_2\,k_3)\,=\,{\pi
\,\beta(\Delta) \over 4\,k_1\, k_2\, k_3}\,
\left ( {k_1\,k_2 \over k_3} \right )^\Ll \, 
(1\,-\,\Delta^2)^{\Ll/2}\, P_{\Ll^\prime}^{-\Ll}(\Delta),
\eqn\THREEPONE$$
when $k_1$, $k_2$ and $k_3$ satisfy the triangular condition
$$ |\,k_1\,-\,k_2\,| \,\le \,k_3 \, \le \,k_1\,+\,k_2.
\eqn\THREEPTWO$$

If we set $\Ll=\Llt=0$ and $\Llo=\Llth=\Ll^\prime$ in eq.~$\TWOPTEN$ and equate
it to eq.~$\THREEPONE$ after setting $\Ll=0$ and interchanging $k_2$
and $k_3$, the following sum rule for the Legendre polynomial is obtained
$$\beta (\eta)\,\sum_{{\cal L}=0}^\Ll\,{\Ll \choose {\cal L}}\,\left (
-\,{k_2\over k_1}\right )^{\cal L}\,P_{\cal L}(\eta)\,=\,
\beta(\eta^\prime)\,\left ({k_3\over k_1}\right )^\Ll\,
P_\Ll(\eta^\prime),\eqn\THREEPTHREE$$
where $\eta\,=\,(k_1^2\,+\,k_2^2\,-k_3^2)/2k_1k_2$ and
$\eta^\prime \,=\,(k_1^2\,+\,k_3^2\,-\,k_2^2)/2k_1k_3$.
Equation $\THREEPTHREE$ is consistent with the result of 
reference [\FIN]. 
In a triangle of sides $k_1$, $k_2$ and $k_3$, $\eta$ is the cosine of the angle
facing side $k_3$ and $\eta^\prime$
is the cosine of the angle facing side $k_2$. 
\chapter{Applications}
One application is an integral over four spherical Bessel functions which arises
in the angular integration of the homogeneous and velocity isotropic Boltzmann
equation [\HOH,\HOHDIS]
$$\eqalign {&I(L\,;\,N+M-L,\,0,\,N,\,M;\,k_1\,k_2\,k_3\,k_4)\,\cr\bspac
&\equiv\,\int_0^\infty \,r^{2-L}\,j_{N+M-L}(k_1r)\,
j_{0}(k_2r)\,j_N(k_3r)\,j_M(k_4r)\,dr,}\eqn\FOURPONE$$
where $L$, $N$ and $M$ are non-negative integers with $N+M\geq L$ 
and $k_1$, $k_2$, $k_3$ and $k_4$ form the sides of 
a quadrilateral.
Using the closure relation for the Bessel functions [2,19], this integral can be
written as
$$\eqalign{&{2\over
\pi}\,\int_0^\infty\,K^2\,dK\,I(L\,;\,N+M-L,\,0,\,N+M-L\,;\,k_1\,k_2\,K)\,
\cr\bspac 
&\times I(0\,;\,N,\,M,\,N+M\,;\,k_3\,k_4\,K).}\eqn\FOURPTWO$$

Now, using eq.~$\TWOPTEN$ 
$$\eqalign{&I(L\,;\,N+M-L,\,0,\,N+M-L\,;\,k_1\,k_2\,K)\,=\,{\pi \, \beta(\xi)
\over
4\,k_1\,k_2\,K}\,\left ( {k_1\over K}\right )^{N+M-L}\cr\bspac
&\times\left ( {k_1\,k_2 \over K } \right )^L\,(1-\xi^2)^{L/2}\,
\sum_{{\cal L}=0}^{N+M-L}\,(-k_2/k_1)^{\cal L}\,{N+M-L \choose {\cal L}}\,P_{\cal
L}^{-L}(\xi),}\eqn\FOURPTHREE$$ 
where $\xi\,=\,(k_1^2+k_2^2-K^2)/2k_1k_2$ and using
$${2l\choose 2l^\prime}^{1/2}\,\threej {l}{l-l^\prime}{l^\prime}
{0}{0}{0}\,=\,{(-1)^l\over \sqrt{2l+1}}\,{l\choose l^\prime}.
\eqn\FOURPFOUR$$
The resulting integral is
$$\eqalign{&I(L\,;\,N+M-L,\,0,\,N,\,M;\,k_1\,k_2\,k_3\,k_4)\,=\,{\pi^2\over 16
k_1k_2k_3k_4}\cr\bspac &\times (k_1\,k_3)^{N+M}\,k_2^{~L}\,\sqrt{2(N+M)+1}\,
{\threej {N}{M}{N+M}{0}{0}{0}}^{-1}\cr\bspac &\times 
\sum_{{\cal L}=0}^{N+M-L}\,\sum_{{\cal L}^\prime=0}^{N+M}\,
(-k_2/k_1)^{\cal L}\,(k_4/k_3)^{{\cal L}^{\,\prime}}\,
{2N+2M\choose 2{\cal L}^\prime}^{1/2}\,{N+M-L \choose {\cal L}}\,
\sum_l\,(2l+1)\cr\bspac &\times 
\threej {N}{N+M-{\cal L}^\prime}{l}{0}{0}{0}\,\threej {M}{\cal
L^\prime}{l}{0}{0}{0}\,\sixj {N}{M}{N+M}{\cal L^\prime}
{N+M-\cal L^\prime}{l}\cr\bspac & \times
S(k_1 k_2 k_3 k_4; L M N {\cal L}^\prime l),}\eqn\FOURPFIVE$$
where, using $\xi^\prime \,=\,(k_3^2+k_4^2-K^2)/2k_3k_4$,
$$S(k_1 k_2 k_3 k_4; L M N {\cal L}^\prime l)\,=\,\int_0^\infty \,
{\beta (\xi) \,\beta(\xi^\prime) \over K^{2(N+M)}}\,(1-\xi^2)^{L/2}\,
P_{\cal L}^{-L}(\xi)\,P_l(\xi^\prime)\,dK,\eqn\FOURPSIX$$
is an integral that is easily done numerically as it converges 
rapidly.

\chapter{Conclusions}
A new generalised analytical formula for the infinite integral over 
three spherical Bessel functions was shown. It involves finite sums
over the associated Legendre function combined with angular momentum 
coupling coefficients 3j and 6j symbols. The sums involved values of the order
which exceeded the degree, which is in conflict with the usual definition of the
associated Legendre functions. An extension for the associated Legendre
functions is shown in appendix A.
\endpage

\APPENDIX{A.}{A: A Generalised Solution to the Associated Legendre Equation}
The associated Legendre function of the first kind, $P_l^m (x)$ is a solution of
the differential equation [\GRADST]
$$(1-x^2){d^2 P_l^m(x) \over dx^2}-2x{dP_l^m(x) \over dx}+[l(l+1)-{m^2 \over
1-x^2}]\,P_l^m(x)=0,\,\eqn\APONE $$
given by
$$P_l^m(x)=(-1)^m\,{(1-x^2)^{m/2} \over 2^l l!}\,{d^{l+m}\over
dx^{l+m}}\,(x^2-1)^l,\,\eqn\APTWO $$
defined for integer degree $l \ge 0$ and integer order $m$ (with $|m| \le l$).
In this appendix, it will be shown that 
another solution exists for this differential equation that extends the
associated Legendre function to 
any rational $|m|$ ($|m| < \infty$) for integer $l$. This solution is irregular
at $x=1$ when $m > l$ and
irregular at $x=-1$ for $m < l$.
The formula can be used to derive closed form expressions for 
$P_l^m$ for a specific $l$ and any rational $m$.

If we set $l=0$ in eq.~$\APONE$, it is easy to show that the following
function, $P_0^m (x)$, is a solution: 
$$P_0^m(x) = a_m \,\Big( {1-x \over 1+x} \Big)^{-m/2}, \eqn\APTHREE$$

where $a_m$ is a coefficient that depends on $m$.
Now, if we assume that the dependence on $l$ can be separated, then in general 
$P_l^m$ can be written as
$$P_l^m(x) = A_{l,m}(x)\, P_0^m(x), \eqn\APFOUR$$
where $A_{l,m}(x)$ is a coefficient that depend on $l$, $m$ and $x$.
To determine the coefficient (up to a constant), substitute eq.~$\APFOUR$
$\,$ back into eq.~$\APONE$.
The resulting differential equation for $A_{l,m}$ is
$$ (1-x^2)\,{d^2 A_{l,m}(x) \over dx^2} + (2m-2x)\,{dA_{l,m}(x) \over dx} +
l(l+1)A_{l,m}(x)=0.\eqn\APFIVE$$
This identifies $A_{l,m}$ as the Jacobi polynomial, [\GRADST]
$P_l^{\,-m,\,\,m}(x)$, defined by
$$P_l^{\,a,\,\,b}(x)={(-1)^l \over 2^l l!}\,(1-x)^{-a}\,(1+x)^{-b}\,{d^l \over
dx^l} [(1-x)^{l+a}\,(1+x)^{l+b}].
\eqn\APSIX $$
Hence, equation \APFOUR $\,$ becomes
$$ P_l^m(x) = b_{l,m}\,\Big({1-x \over 1+x}\Big)^{-m/2}\,P_l^{\,-m,\,\,m}(x),
\eqn\APSEVEN $$
where $b_{l,m}$ is a constant that depends on $l$ and $m$. 
Using the normalisation for the associated Legendre functions
$$\int_{-1}^1\,[P_l^m(x)]^2 dx = {2\over
2l+1}\,{(l+m)!\over(l-m)!},\eqn\APEIGHT$$
and the normalisation for the Jacobi polynomials
$$\int_{-1}^1\,(1-x)^a\,(1+x)^b\,[P_l^{\,a,\,\,b}(x)]^2 dx =
2^{a+b+1}\,{(l+a)!(l+b)!\over 
l!(a+b+2l+1)(l+a+b)!},\eqn\APNINE$$
one can square eq.~$\APSEVEN$ and integrate it over $x$ from $-1$ to $1$,
then compare it with 
eq.~$\APEIGHT$ to obtain an expression for $b_{l,m}$:
$$b_{l,m}= {l!\over |l-m|!},\eqn\APTEN $$
where the absolute value in $|l-m|!$ is introduced to allow m to go from
$-\infty$ to $\infty$, 
and the expression reduces to $(l-m)!$ for $-l \le m \le l$. 
Equation $\APSEVEN$ then becomes
$$P_l^m(x)={(-1)^l\over 2^l |l-m|!}\,\Big({1-x \over 1+x}\Big)^{m/2}\,{d^l \over
dx^l}\,[(1-x^2)^l\Big({1+x\over 1-x}\Big)^m],
\eqn\APELEVEN$$
which agrees with eq.~(2.4), page 12 of reference [\VIR] and is valid for
integer degree 
$l\ge 0$ and rational order $-\infty < m < \infty$.

When $-l \le m \le l$, equations $\APELEVEN$ and $\APTWO$ are equal, 
leading to the identity
$$ {d^{l+m} \over dx^{l+m}}\,(1-x^2)^l\,=\,(-1)^m\,{l!\over
(l-m)!}\,(1+x)^{-m}\,{d^l\over dx^l}\,[(1-x^2)^l\,
\Big({1+x\over 1-x}\Big)^m],\eqn\APTWELVE$$
for $-l \le m \le l$.

Another property that can easily be found is
$$P_l^m(-x)\,=\,(-1)^l\,{|l+m|!\over |l-m|!}\,P_l^{-m}(x),\eqn\APTHIRTEEN$$
for positive integer $l$ and any rational $m$.

Using equations $\APSEVEN$ and $\APTEN$, one can express the associated Legendre
function in terms of the Jacobi 
polynomial as
$$P_l^m(x)\,=\,{l!\over |l-m|!}\,\Big({1-x\over
1+x}\Big)^{-m/2}\,P_l^{\,-m,\,\,m}(x).\eqn\APFOURTEEN$$
Using the recurrence relation for the Jacobi polynomials [\GRADST] 
$$\eqalign{&2(n+1)\,(n+\alpha +\beta +1)(2n+\alpha
+\beta)\,P_{n+1}^{(\alpha,\,\beta)}(x)\,=\,(2n+\alpha+\beta+1)\cr\bspac
&\times [(2n+\alpha+\beta)(2n+\alpha+\beta
+2)\,x\,+\alpha^2\,-\,\beta^2]\,P_n^{(\alpha,\,\beta)}(x)\,-\,2(n+\alpha)
(n+\beta)\cr\bspac
&\times(2n+\alpha+\beta+2)P_{n-1}^{(\alpha,\,\beta)}(x),}\eqn\APFIFTEEN$$
we find the following recurrence relation for the generalised associated
Legendre functions
$$|\,l-m+1\,|\,P_{l+1}^m(x)\,=\,(2l+1)\,x\,P_l^m(x)-{(l^2-m^2)\over
|\,l-m\,|}\,P_{l-1}^m(x).\eqn\APSIXTEEN$$
Using this recurrence relation and eq.~$\APELEVEN$ one finds the following
closed-form expressions for the associated Legendre functions at specific $l$'s and any rational m
$$P_0^m(x)\,=\,{1\over |\,m\,|\,!}\,\,\Big({1+x\over 1-x}\Big)^{m/2},\eqn\APSEVENTEEN$$
$$P_1^m(x)\,=\,\,{1\over|\,1-m\,|\,!}\,\,(x-m)\,\,\Big({1+x\over1-x}\Big)^{m/2},
\eqn\APEIGHTEEN$$
$$P_2^m(x)\,=\,{1\over |\,2-m\,|\,!}\,\,(3\,{x}^{2}-3\,xm-1+{m}^{2})\,\Big({1+x\over
1-x}\Big)^{m/2},\eqn\APNINETEEN$$
$$P_3^m(x)\,=\,{1\over
|\,3-m\,|\,!}\,\,(15\,{x}^{3}-15\,{x}^{2}m-9\,x+6\,x{m}^{2}+4\,m-{m}^{3})\,\Big({
1+x\over 1-x}\Big)^{m/2},
\eqn\APTWENTY$$
$$\eqalign{P_4^m(x)\,=\,{1\over
|\,4-m\,|\,!}\,\,(&105\,{x}^{4}-105\,{x}^{3}m-90\,{x}^{2}+45\,{x}^{2}{m}^{2}+55\,
xm\cr
\bspac
&-10x{m}^{3}+9-10\,{m}^{2}+{m}^{4})\,\Big({1+x\over 1-x}\Big)^{m/2}.}\eqn\APTWENTYONE$$

\def\frac#1#2{{{#1}\over {#2}}}
\def\sgn{sgn}
\APPENDIX {B.}{B: Explicit Expressions}
Here we provide some explicit expressions for the integral $\ONEPTWO$ and compare to those obtained from a result given in [\GERVA], which reads in our notation
$$\eqalign{I(\lambda;\,\lambda_1,\,\lambda_2,\,\lambda_3\,;\,k_1\,k_2\,k_3)=&-\pi \,{i}^{{\lambda_1}+{\lambda_2}+{\lambda_3}} \cr
\bspac
\times\Big(& \sgn(k_1+k_3+k_2)F_{\lambda\,\,\lambda_1\,\lambda_2\,\lambda_3}^{m_1\,m_2\,m_3}(k_1,\,k_2,\,k_3)\cr
\bspac
+ \left( -1 \right) ^{{\lambda_1}}&\sgn(-k_1+k_2+k_3)F_{\lambda\,\,\lambda_1\,\lambda_2\,\lambda_3}^{m_1\,m_2\,m_3}(-k_1,\,k_2,\,k_3)\cr
\bspac
+ \left( -1 \right) ^{{\lambda_2}}&\sgn(k_1-k_2+k_3)F_{\lambda\,\,\lambda_1\,\lambda_2\,\lambda_3}^{m_1\,m_2\,m_3}(k_1,\,-k_2,\,k_3)\cr
\bspac
+ \left( -1 \right) ^{{\lambda_1}+{\lambda_2}}&\sgn(-k_1-k_2+k_3)F_{\lambda\,\,\lambda_1\,\lambda_2\,\lambda_3}^{m_1\,m_2\,m_3}(-k_1,\,-k_2,\,k_3) \Big) .}\eqn\APBONE$$
with 
$$\eqalign{F_{\lambda\,\,\lambda_1\,\lambda_2\,\lambda_3}^{m_1\,m_2\,m_3}(k_1,\,k_2,\,k_3)&=\sum _{{m_1}=0}^{{\lambda_1}} \frac{ \left( {\lambda_1}+{m_1} \right) !\, \left( -1 \right) ^{{m_1}} }{\left( {\lambda_1}-{m_1} \right) ! {m_1}!  \left( 2\,{k_1} \right) ^{{m_1}+1} }\sum _{{m_2}=0}^{{\lambda_2}} \frac{ \left( {\lambda_2}+{m_2} \right) !\, \left( -1 \right) ^{{m_2}} }{\left( {\lambda_2}-{m_2} \right) ! {m_2}! \left( 2\,{k_2} \right) ^{{m_2}+1} }\cr
\bspac
\times&\sum _{{m_3}=0}^{{\lambda_3}+{\lambda}}{\frac { \left( {\lambda_3}+{\lambda}+{m_3} \right) !\, \left( -1 \right) ^{{m_3}} }{ \left( {\lambda_3}+{\lambda}-{m_3} \right) !\,{m_3}!\, \left( 2\,{k_3} \right) ^{{m_3}+1} } \frac{\left( {k_1}+{k_2}+{k_3} \right) ^{{m_1}+{m_2}+{m_3}+{\lambda}}}{\left( {m_1}+{m_2}+{m_3}+{\lambda} \right) !}}. }\eqn\APBTWO$$
Restricting ourselves to values of $k_1$, $k_2$ and $k_3$ which satisfy the triangular condition and using symmetries of $F_{\lambda\,\,\lambda_1\,\lambda_2\,\lambda_3}^{m_1\,m_2\,m_3}$, $\APBONE$ can be simplified to 
$$\eqalign{ I(\lambda;\lambda_1,\lambda_2,\lambda_3;&k_1,k_2,k_3)=
 -\pi \,\beta \left( \Delta \right) {i}^{ \lambda_1+ \lambda_2+ \lambda_3}\cr
\bspac
\times&\sum _{ m_1=0}^{ \lambda_1}   \frac{\left( -1 \right) ^{ m_1}\left(  \lambda_1+ m_1 \right) !\, }{\left(  \lambda_1- m_1 \right)!  m_1!  \left( 2\,k_1 \right) ^{ m_1+1}}\sum _{ m_2=0}^{ \lambda_2}   \frac{\left( -1 \right) ^{ m_2}\left(  \lambda_2+ m_2 \right) !}{  \left(  \lambda_2- m_2 \right)! m_2! \left( 2\,k_2 \right) ^{ m_2+1}} \cr
\bspac
&\sum _{ m_3=0}^{ \lambda_3+\lambda}{\frac {\left( -1 \right)^{ m_3}\left(  \lambda_3+\lambda+ m_3 \right)!}{ \left(  \lambda_3+\lambda- m_3 \right) !\, m_3!\, \left( 2\,k_3 \right) ^{ m_3+1}}\frac{1}{\left(  m_1+ m_2+ m_3+\lambda \right) !}}\cr
\bspac
& \times \bigg[ \left( k_1+k_2+k_3 \right) ^{ m_1+ m_2+ m_3+\lambda}- \left( -1\right) ^{ \lambda_1+ m_1}{c_1}^{ m_1+ m_2+ m_3+\lambda}\cr
\bspac
&- \left( -1 \right) ^{ \lambda_2+ m_2}{c_2}^{ m_1+ m_2+ m_3+\lambda}- \left( -1 \right) ^{ \lambda_3+\lambda+ m_3}{c_3}^{ m_1+ m_2+ m_3+\lambda} \bigg]},\eqn\APBTHREE$$
with $c_1=\left(-k_1+k_2+k_3\right)$, $c_2=\left(k_1-k_2+k_3\right)$ and $c_3=\left(k_1+k_2-k_3\right)$. For numerical computations this expression is more suitable than $\APBONE$ in which almost cancellations between the different terms including $\sgn$ functions can lead to large relative errors.
For some small values of $\lambda$, $\lambda_1$, $\lambda_2$ and $\lambda_3$ this yields the following explicit results:
$$\eqalign{&I(0;0,0,0;k_1k_2k_3)=\frac{1}{4}{\frac {\pi \beta ( \Delta ) }{{k_1}{k_2}{k_3}}},}$$
$$\eqalign{I(0;1,0,1;k_1k_2k_3)= -\frac{1}{16}\frac {\pi \beta ( \Delta ) }{{{k_1}}^{2}{k_2}{{k_3}}^{2}} 
\Big(& -2{k_1}{k_3}+{{k_1}}^{2}+2{k_1}{c_1}-2{k_1}{c_2}-2{k_1}{c_3}+{{k_3}}^{2}\cr
\bspac
&-2{k_3}{c_1}-2{k_3}{c_2}+2{k_3}{c_3}-{{k_2}}^{2}+{{c_1}}^{2}+{{c_2}}^{2}+{{c_3}}^{2} \Big) ,}$$
$$\eqalign{I(0;1,1,0;k_1k_2k_3)=&-\frac{1}{16}\frac {\pi \beta ( \Delta ) }{{{k_1}}^{2}{{k_2}}^{2}{k_3}} \Big( -2{k_1}{k_2}+{{k_1}}^{2}+2{k_1}{c_1}-2{k_1}{c_2}-2{k_1}{c_3}\cr
\bspac
&+{{k_2}}^{2}-2{c_1}{k_2}+2{k_2}{c_2}-2{k_2}{c_3}-{{k_3}}^{2}+{{c_1}}^{2}+{{c_2}}^{2}+{{c_3}}^{2} \Big) ,}$$
$$\eqalign{I(1;0,0,0;k_1k_2k_3)=-\frac{1}{16}\frac {\pi \beta ( \Delta ) }{{k_1}{k_2}{{k_3}}^{2}} \Big(& {{k_3}}^{2}-2{k_3}{c_1}-2{k_3}{c_2}+2{k_3}{c_3}-{{k_1}}^{2}-2{k_1}{k_2}\cr
\bspac
&-{{k_2}}^{2}+{{c_1}}^{2}+{{c_2}}^{2}+{{c_3}}^{2} \Big) ,}$$
$$\eqalign{&I(1;1,0,1;k_1k_2k_3)={\frac {1}{64}}\frac {\pi \beta ( \Delta ) }{{{k_1}}^{2}{k_2}{{k_3}}^{3}} \Big( 2{{k_2}}^{2}{{k_3}}^{2}+4{{k_3}}^{2}{{c_1}}^{2}-12{k_1}{{c_1}}^{2}{k_3}+{{c_2}}^{4}+8{k_1}{{k_3}}^{2}{c_1}\cr
\bspac
&-8{k_1}{{k_3}}^{2}{c_3}-8{k_1}{{k_3}}^{2}{c_2}+12{k_1}{k_3}{{c_2}}^{2}-12{k_1}{k_3}{{c_3}}^{2}+6{{k_1}}^{2}{{k_2}}^{2}-4{k_1}{{c_2}}^{3}-4{k_1}{{c_3}}^{3}+4{k_1}{{c_1}}^{3}\cr
\bspac
&+4{k_3}{{c_3}}^{3}+4{{k_3}}^{2}{{c_2}}^{2}+4{{k_3}}^{2}{{c_3}}^{2}-4{k_3}{{c_2}}^{3}-4{k_3}{{c_1}}^{3}-2{{k_1}}^{2}{{k_3}}^{2}+8{{k_1}}^{3}{k_2}+{{c_1}}^{4}\cr
\bspac
&+{{c_3}}^{4}+3{{k_1}}^{4}-{{k_2}}^{4}-{{k_3}}^{4} \Big) ,}$$
$$\eqalign{&I(1;1,1,0;k_1k_2k_3)={\frac {1}{192}}\frac {\pi \beta ( \Delta ) }{{{k_1}}^{2}{{k_2}}^{2}{{k_3}}^{2}} \Big( -4{k_2}{{c_1}}^{3}-6{{k_2}}^{2}{{k_3}}^{2}+24{k_1}{k_2}{k_3}{c_2}-12{k_1}{{c_1}}^{2}{k_3}\cr
\bspac
&+{{c_2}}^{4}+12{k_1}{k_3}{{c_2}}^{2}-12{k_1}{k_3}{{c_3}}^{2}-12{k_1}{{c_1}}^{2}{k_2}-12{k_2}{k_3}{{c_2}}^{2}+12{k_2}{k_3}{{c_1}}^{2}-12{k_2}{k_3}{{c_3}}^{2}\cr
\bspac
&+12{k_1}{k_2}{{c_3}}^{2}-12{k_1}{k_2}{{c_2}}^{2}+24{k_1}{k_2}{k_3}{c_3}+24{k_1}{k_2}{k_3}{c_1}-4{k_2}{{c_3}}^{3}-6{{k_1}}^{2}{{k_2}}^{2}-4{k_1}{{c_2}}^{3}\cr
\bspac
&-4{k_1}{{c_3}}^{3}+4{k_1}{{c_1}}^{3}+4{k_3}{{c_3}}^{3}-4{k_3}{{c_2}}^{3}-4{k_3}{{c_1}}^{3}-6{{k_1}}^{2}{{k_3}}^{2}+4{k_2}{{c_2}}^{3}+{{c_1}}^{4}+{{c_3}}^{4}\cr
\bspac
&+3{{k_1}}^{4}+3{{k_2}}^{4}+3{{k_3}}^{4} \Big) .} \eqn\APBFOUR$$
Higher order expressions are not shown here for the sake of brevity.
On the other hand we get from eq.~$\TWOPTEN$:

$$\eqalign{I(\lambda\,;\,\lambda_1,\,\lambda_2,\,\lambda_3\,;\,k_1\,k_2\,k_3)=&\frac{\pi\beta(\Delta)}{2^{\lambda+2}(\lambda_1+\lambda_2+\lambda_3+\lambda)!}\cr
\bspac
&\times {k_1}^{-(\lambda_1+1)} {k_2}^{-(\lambda_2+1)} {k_3}^{-(\lambda_3+\lambda+1)} A(\lambda;\lambda_1,\lambda_2,\lambda_3),}\eqn\APBFIVE$$
with
$$A(0;0,0,0)={
1
}$$
$$A(0;1,0,1)={
-2\,k_1  \left( -k_1 +k_2 \Delta  \right) ,
}$$
$$A(0;1,1,0)={
2\,k_1 k_2 \Delta ,
}$$
$$A(1;0,0,0)={
2\,k_1 k_2  \left( 1-\Delta  \right) ,
}$$
$$A(1;1,0,1)={
-6\,{k_1 }^{2}k_2  \left( -2\,k_1 +k_2 \Delta +k_2  \right)  \left( 1-\Delta  \right) ,
}$$
$$A(1;1,1,0)={
6\,{k_1 }^{2}{k_2 }^{2} \left( \Delta +1 \right)  \left( 1-\Delta  \right) ,
}$$
$$A(2;0,0,0)={
4\, \left( \Delta -1 \right) ^{2}{k_2 }^{2}{k_1 }^{2} ,
}$$
$$A(2;1,0,1)={
-16\, \left( \Delta -1 \right) ^{2} \left( -3\,k_1 +2\,k_2 +k_2 \Delta  \right) {k_2 }^{2}{k_1 }^
{3} ,
}$$
$$A(2;1,1,0)={
16\, \left( \Delta -1 \right) ^{2} \left( 2+\Delta  \right) {k_2 }^{3}{k_1 }^{3} .
}$$
To test our new result we have compared several explicit analytic expressions obtained from $\TWOPTEN$ and $\APBTHREE$ respectively (including the ones given above), and found full agreement. The advantage of equations $\APBFIVE$ over $\APBFOUR$ is their compact form in terms of generalized Legendre functions. This property will be exploited in [\INPREP] where we study the integral 
$$\int_0^\infty\,e^{-r/u}r^{n}\,j_{\Llo}(k_1r)\,j_{\Llt}(k_2r)\,
dr.\eqn\APBSIX$$ 

\endpage

\par \penalty-400 \vskip\chapterskip
   \spacecheck\referenceminspace \immediate\closeout\referencewrite
   \referenceopenfalse
   \line{\fourteenrm\hfil References\hfil}\vskip\headskip
   \input referenc.tex

\endpage
\bye